\documentclass[%
  reprint,
 amsmath,amssymb,
 aps,
floatfix,
]{revtex4-1}
\usepackage[T1]{fontenc}
\usepackage{float}
\usepackage{empheq}
\usepackage{hyperref}
\usepackage{cleveref}
\usepackage{tikz}
\usepackage{cancel}
\usepackage{caption}
\usepackage{sidecap}
\usepackage[colorinlistoftodos,prependcaption]{todonotes}
\usepackage{comment}
\newcommand{\be}{\begin{equation}}
\newcommand{\ee}{\end{equation}}

\newcommand{\bea}{\begin{eqnarray}}
\newcommand{\eea}{\end{eqnarray}}

\begin{document}
\title{Gravitational Sedimentation and Rebound of Strongly Coupled Dusty Plasma Crystals: A Molecular Dynamics Study}

\author{Gurudatt Gaur}
\thanks{Corresponding author: gurudatt.gaur@sxca.edu.in}
\affiliation{St. Xavier’s College (Autonomous), Navrangpura, Ahmedabad, GJ, 380 009, India}
\author{Meet Contractor}
\email{contractormeet04@gmail.com}
\affiliation{St. Xavier’s College (Autonomous), Navrangpura, Ahmedabad, GJ, 380 009, India}
\author{Vikram Dharodi}
\email{vikram.ipr@gmail.com}
\affiliation{Department of Physics and Astronomy, West Virginia University, Morgantown, WV, 26506, USA}

\date{\today}

\begin{abstract}
The gravitational sedimentation of strongly coupled dusty plasma crystals is investigated using molecular dynamics simulations. Initially, the dust particles are levitated by the balance between the upward external electric field and gravity. Sedimentation is initiated by removing the electric field, allowing the particles to settle collectively under gravity while interacting through the Yukawa (screened Coulomb) potential. Single-layer, AB-stacked bilayer, and ABA-stacked trilayer crystals are investigated to examine the influence of crystal geometry on the sedimentation dynamics. All crystal configurations undergo collective gravitational settling while preserving their in-plane hexagonal ordering during the initial stages of sedimentation. Upon collision with a reflecting boundary, the multilayer crystals undergo transient interlayer compression followed by sequential momentum transfer between neighboring layers, producing coherent collective rebound. In particular, the trilayer exhibits sequential layer-by-layer momentum propagation from the lower to the middle and finally to the upper layer. During successive sedimentation--rebound cycles, repeated interlayer interactions progressively degrade the initial ABA stacking while preserving the collective mechanical response of the crystal. These results demonstrate that strong Yukawa coupling enables multilayer dusty plasma crystals to sustain repeated impacts while maintaining coherent collective motion despite gradual structural evolution. The present study provides a particle-resolved description of gravitational sedimentation in multilayer dusty plasma crystals and offers a theoretical framework for interpreting laboratory experiments following the removal of electrostatic confinement.
\end{abstract}

\maketitle

\section{Introduction}\label{Introduction}

Dusty plasmas are multicomponent plasmas consisting of electrons, ions, neutral particles, and charged dust grains. Owing to the much higher mobility of electrons than ions, dust particles generally acquire a large negative charge through the collection of plasma electrons and ions. Under certain conditions, however, such as secondary electron emission, dust particles may instead become positively charged~\cite{horanyi1996charged,ishihara2007complex,chaubey2021positive}. Depending on the plasma conditions, micron-sized dust particles typically carry charges ranging from $10^3$ to $10^5$ elementary charges while possessing masses several orders of magnitude larger than those of electrons and ions. Their large mass and charge give rise to a variety of collective phenomena that are absent in conventional electron--ion plasmas, making dusty plasmas an important platform for studying strongly coupled many-body systems.

Dusty plasmas occur in a broad range of laboratory, industrial, and astrophysical environments~\cite{merlino2004dusty,adamovich20222022}. In plasma-assisted semiconductor manufacturing, plasma etching, plasma-enhanced chemical vapor deposition, and magnetic confinement fusion devices, dust particles are generally regarded as undesirable contaminants because they can become trapped in plasma sheaths and eventually deposit on wafer or chamber surfaces, thereby reducing manufacturing yield~\cite{bouchoule1999dusty,federici2001plasma,fortov2005complex,ramkorun2024introducing,lee2022low}. In contrast, dust plays a fundamental role in astrophysical environments, contributing to the formation and evolution of planets, moons, asteroids, and other celestial bodies~\cite{natta2006dust,shukla2015introduction,armitage2020astrophysics}. Consequently, understanding the transport, dynamics, and sedimentation of charged dust is important in both technological applications and naturally occurring dusty plasma systems.

In laboratory experiments, dust particles are deliberately introduced into plasmas to create well-controlled model systems for investigating strongly coupled many-body physics. Under suitable plasma conditions, the electrostatic interaction between highly charged dust particles becomes sufficiently strong to overcome thermal motion, resulting in the formation of ordered plasma crystals~\cite{thomas1994plasma,singh2022square}. These crystals provide an ideal platform for studying lattice waves, phase transitions, melting, transport, and nonlinear excitations with single-particle resolution~\cite{nunomura2000transverse,thomas2016initial,choudhary2016transport,piel2017plasma,choudhary2020influence,kumar2022trapping,dharodi2023ring}. In most laboratory experiments, dust particles are levitated above the lower electrode by the balance between the upward sheath electric force and gravity. When the confining electric field is removed, the particles lose their electrostatic support and undergo gravitational sedimentation while continuing to interact through the screened Coulomb (Yukawa) potential.

Gravitational sedimentation provides a simple yet fundamental process for investigating the interplay between gravity and strong interparticle interactions. Recent afterglow-plasma experiments by Chaubey \textit{et al.}~\cite{chaubey2021positive,chaubey2022preservation,chaubey2023controlling} demonstrated that dust crystals undergo collective sedimentation following the removal of electrostatic confinement, revealing rich particle dynamics during free fall. These experiments have stimulated renewed interest in understanding the collective mechanical response of strongly coupled dusty plasma crystals during sedimentation.

Despite these experimental advances, several fundamental questions remain unanswered. In particular, the influence of crystal geometry on collective sedimentation, rebound dynamics, interlayer momentum transfer, and structural evolution has not been systematically investigated. It remains unclear how multilayer plasma crystals respond following impact with a confining boundary, how momentum propagates between neighboring layers, and whether repeated sedimentation--rebound cycles preserve or progressively degrade the initial crystalline ordering.

In the present work, we employ molecular dynamics simulations to investigate the gravitational sedimentation of single-layer, AB-stacked bilayer, and ABA-stacked trilayer dusty plasma crystals. An idealized model is considered in which the electrostatic levitation is removed while the dust charge and Yukawa interaction remain constant, thereby isolating the collective mechanical response of strongly coupled dust crystals. Unlike previous studies that primarily focused on the settling process itself, we demonstrate that multilayer crystals exhibit rich rebound dynamics governed by strong Yukawa coupling. In particular, impact with a reflecting boundary produces transient interlayer compression followed by sequential momentum transfer between neighboring layers. Repeated sedimentation--rebound cycles progressively degrade the initial ABA stacking while preserving coherent collective motion. These findings reveal previously unexplored aspects of the mechanical response of multilayer dusty plasma crystals and establish a particle-resolved framework for understanding impact dynamics in strongly coupled dusty plasmas.

Molecular dynamics (MD) simulations provide a natural framework for investigating these processes because they resolve the motion of individual particles while fully accounting for many-body Yukawa interactions. Unlike continuum descriptions, MD simulations enable direct visualization of particle trajectories, structural evolution, momentum transfer, and collective dynamics throughout the sedimentation and rebound processes, thereby providing particle-level insight into the underlying physical mechanisms.

Although the present study is motivated by fundamental dusty plasma physics, the underlying physical mechanisms are also relevant to other plasma environments. In particular, charged particulate contaminants in plasma-assisted semiconductor manufacturing may become trapped within plasma sheaths and subsequently settle onto wafer or chamber surfaces following changes in plasma operating conditions. While the present model is idealized, it provides fundamental insight into the collective transport, momentum transfer, and impact dynamics of charged particles in plasma environments.

The remainder of this paper is organized as follows. Section~\ref{sec:model} describes the numerical model and simulation methodology. Section~\ref{Results_discussion} presents the simulation results and discusses the sedimentation and rebound dynamics of single-layer, bilayer, and trilayer dusty plasma crystals. Finally, Section~\ref{Conclusions} summarizes the main conclusions and outlines directions for future work.

\section{Numerical Model}
\label{sec:model}

The gravitational sedimentation of strongly coupled dusty plasma crystals is investigated using the open-source molecular dynamics package LAMMPS~\cite{plimpton1995fast}. The simulations are performed in a two-dimensional computational domain with periodic boundary conditions in the horizontal ($x$) direction and confined boundaries in the vertical ($y$) direction. The computational domain has dimensions $L_x=L_y=20a$, where $a=0.05~\mathrm{cm}$ is the characteristic interparticle spacing, corresponding to a physical domain of $1.0\times1.0~\mathrm{cm}^2$. A small finite thickness is assigned in the out-of-plane direction to preserve the two-dimensional geometry.

The dust particles are modeled as identical charged particles with mass
$m_d=6.99\times10^{-10}~\mathrm{g}$ and charge $q=-11940e$, where $e$ is the elementary charge. The interaction potential energy between two dust particles is described by the Yukawa potential,

\begin{equation}
U^{\rm Yuk}(r)=\frac{q^2}{r}
\exp\left(-\frac{r}{\lambda_D}\right),
\end{equation}

where $r$ is the interparticle distance and $\lambda_D$ is the Debye screening length. The screening parameter is fixed at $\kappa=a/\lambda_D=0.5$, and the interaction is truncated at a cutoff distance of $r_c=5a$.

The Coulomb coupling parameter,
\begin{equation}
\Gamma=\frac{q^2}{ak_BT},
\end{equation}
is prescribed as $\Gamma=2000$, from which the initial temperature is determined. Additional simulations performed over the range $200 \le \Gamma \le 2000$ exhibited qualitatively similar dynamics; therefore, only the representative $\Gamma=2000$ results are presented. The initial particle velocities are then sampled from the corresponding Maxwellian distribution.

Initially, the particles are randomly distributed within the central region of the computational domain and subsequently equilibrated to form stable crystalline structures. By varying the number of particles, equilibrium configurations consisting of a single layer, an AB-stacked bilayer, and an ABA-stacked trilayer are obtained.

During equilibration, a vertically varying sheath electric field provides an upward electrostatic force,

\begin{equation}
\mathbf{F}_E=q\mathbf{E},
\end{equation}

where the field is chosen such that the electrostatic force balances gravity at the equilibrium levitation height,

\begin{equation}
qE(y_{\rm eq})=m_dg.
\end{equation}

Under this condition, the dust particles remain levitated and spontaneously self-organize into stable crystalline structures.

The equations of motion are integrated using the velocity-Verlet algorithm. A Nos\'e--Hoover thermostat is applied for $1000\,\omega_{pd}^{-1}$ to maintain the target temperature corresponding to $\Gamma=2000$, where $\omega_{pd}=\sqrt{2q^2/(m_da^3)}$ is the nominal dust plasma frequency. The integration time step is chosen as $\Delta t=5\times10^{-3}\omega_{pd}^{-1}$, which provides adequate temporal resolution while maintaining numerical stability.

After equilibration, the levitating sheath electric field is removed while gravity is retained, initiating gravitational sedimentation. The particles subsequently evolve in the microcanonical ($NVE$) ensemble under the combined influence of gravity and Yukawa interactions.

Reflecting walls are imposed at the upper and lower boundaries of the computational domain to investigate the collision dynamics. Upon impact with the lower boundary, the crystals undergo elastic reflection. For multilayer crystals, the collision produces transient interlayer compression followed by momentum transfer between neighboring layers, causing the crystal to rebound collectively. Throughout the simulations, particle positions, velocities, accelerations, and structural configurations are recorded to characterize the sedimentation, collision, rebound, and structural evolution during repeated oscillation cycles.

The vertical center-of-mass (COM) position of each layer is calculated as
\begin{equation}
y_{\rm COM}(t)=\frac{1}{N}\sum_{i=1}^{N} y_i(t),
\end{equation}
where $N$ is the number of particles in the layer and $y_i(t)$ is the vertical position of particle $i$ at time $t$. The corresponding COM velocity is obtained as
\begin{equation}
v_{\rm COM}(t)=\frac{1}{N}\sum_{i=1}^{N} v_{y,i}(t),
\end{equation}
where $v_{y,i}(t)$ is the vertical velocity of particle $i$.


\section{Results and Discussion}
\label{Results_discussion}

The gravitational sedimentation and rebound dynamics of single-layer, AB-stacked bilayer, and ABA-stacked trilayer dusty plasma crystals are presented in this section. Particular attention is given to the influence of crystal geometry on the collective dynamics, interlayer momentum transfer, and structural evolution during repeated collisions with the reflecting boundary.

\begin{figure}[ht]
\centering
\includegraphics[width=\linewidth]{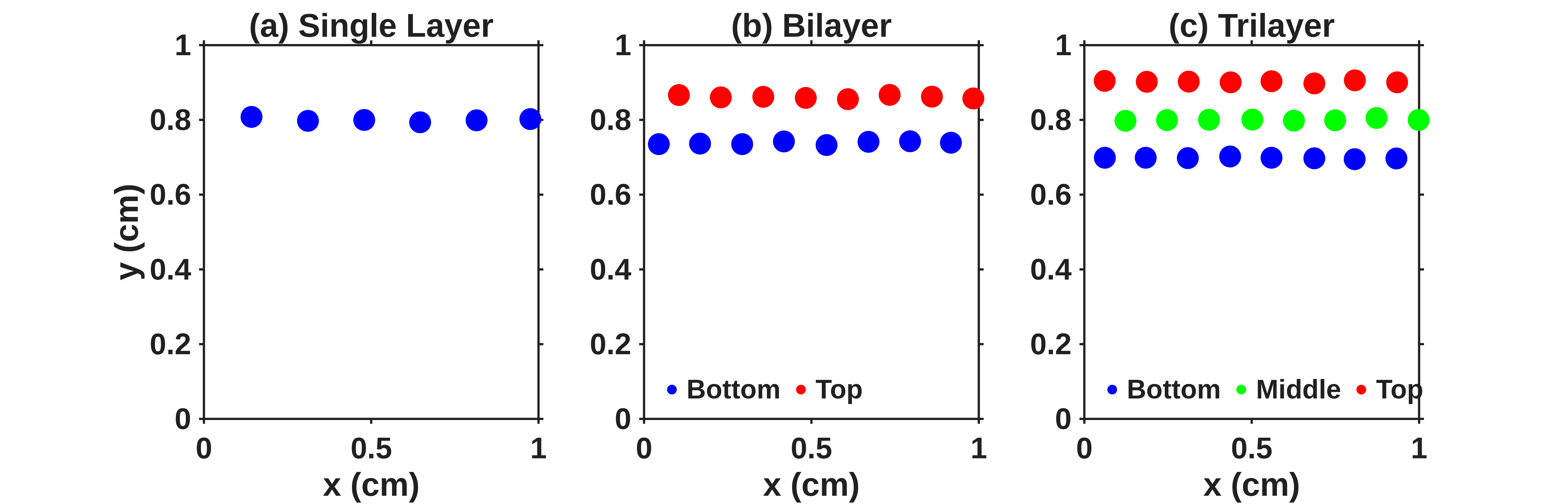}
\caption{
Initial equilibrium configurations used in the molecular dynamics simulations: (a) single-layer, (b) AB-stacked bilayer, and (c) ABA-stacked trilayer dusty plasma crystals. Blue, green, and red particles denote the bottom, middle, and top layers, respectively.
}
\label{fig:figure1}
\end{figure}

Figure~\ref{fig:figure1} presents the initial equilibrium configurations used in the molecular dynamics simulations. All crystal configurations exhibit well-ordered in-plane hexagonal lattices prior to sedimentation. The multilayer crystals are constructed by stacking adjacent layers in AB and ABA sequences for the bilayer and trilayer systems, respectively. These equilibrium structures provide the reference state from which gravitational sedimentation is initiated following the removal of the confining electric field.

\subsection{Sedimentation of a Single Dust Layer}

The sedimentation dynamics of the single-layer dusty plasma crystal are first examined. As shown in Fig.~\ref{fig:figure1}(a), the initial equilibrium configuration consists of a single hexagonally ordered layer levitated at an equilibrium height of approximately $y=0.8~\mathrm{cm}$ by the balance between the upward confining electric force and gravity. Following the removal of the confining electric field at $t=0$, the crystal undergoes collective gravitational sedimentation.

\begin{figure}[ht]
\centering
\includegraphics[width=\linewidth]{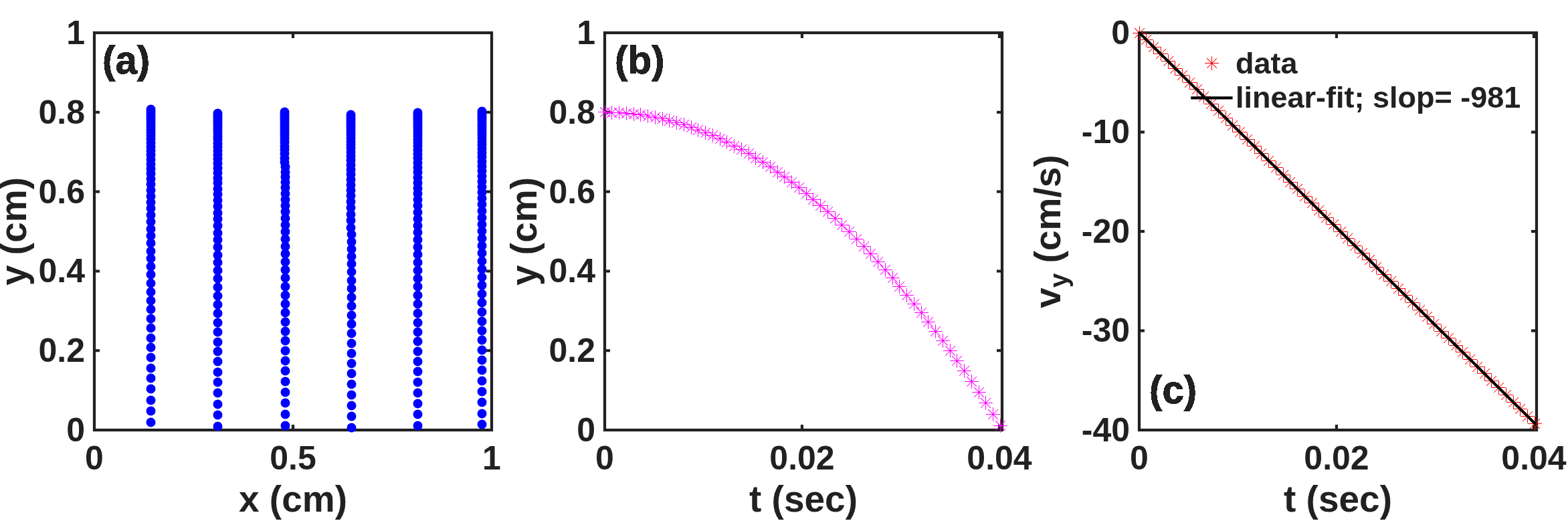}
\caption{
Evolution of the gravitational sedimentation of a single-layer strongly coupled dust crystal following the removal of the external sheath electric field.
(a) Representative particle trajectories.
(b) Vertical particle position as a function of time.
(c) Vertical particle velocity as a function of time. The solid line represents a linear fit with slope $-981~\mathrm{cm\,s^{-2}}$, corresponding to the imposed gravitational acceleration.
}
\label{fig:figure2}
\end{figure}

Figure~\ref{fig:figure2}(a) shows the trajectories of six representative dust particles during the sedimentation process. The particles descend together while maintaining their relative horizontal positions and nearly constant interparticle spacing, indicating that the in-plane crystalline ordering is preserved throughout the motion. No appreciable lateral displacement, structural distortion, or differential settling is observed, demonstrating that the strong Yukawa interactions maintain the structural integrity of the crystal during free fall.

The temporal evolution of the vertical position of a representative dust particle is presented in Fig.~\ref{fig:figure2}(b). The particle height decreases parabolically with time, consistent with motion under constant gravitational acceleration. Starting from the initial equilibrium position, the particle reaches the lower reflecting boundary after approximately $0.040~\mathrm{s}$, in excellent agreement with the analytical free-fall time. Upon collision with the reflecting wall, the particle reverses its direction of motion and rebounds upward. The upward motion is subsequently decelerated by gravity until the particle reaches its maximum height, after which the next sedimentation cycle begins.

The corresponding vertical velocity is shown in Fig.~\ref{fig:figure2}(c). During the free-fall stage, the velocity varies linearly with time, and a linear fit yields a slope of $-981~\mathrm{cm\,s^{-2}}$, identical to the imposed gravitational acceleration. The abrupt change in the velocity at the reflecting boundary corresponds to the elastic collision with the wall, after which gravity continuously reduces the upward velocity until it reaches zero at the turning point. The excellent agreement between the numerical results and the analytical prediction confirms that the particles undergo nearly ideal free-fall motion without appreciable numerical drift or artificial damping following the removal of the confining electric field.

Although each particle experiences gravitational acceleration, the strong Yukawa coupling continuously redistributes momentum among neighboring particles, suppressing relative motion and preserving the crystalline arrangement throughout both the sedimentation and rebound processes. Consequently, the dust crystal behaves as a coherent strongly coupled many-body system rather than as a collection of independently falling particles.

\begin{figure}[ht]
\centering
\includegraphics[width=\linewidth]{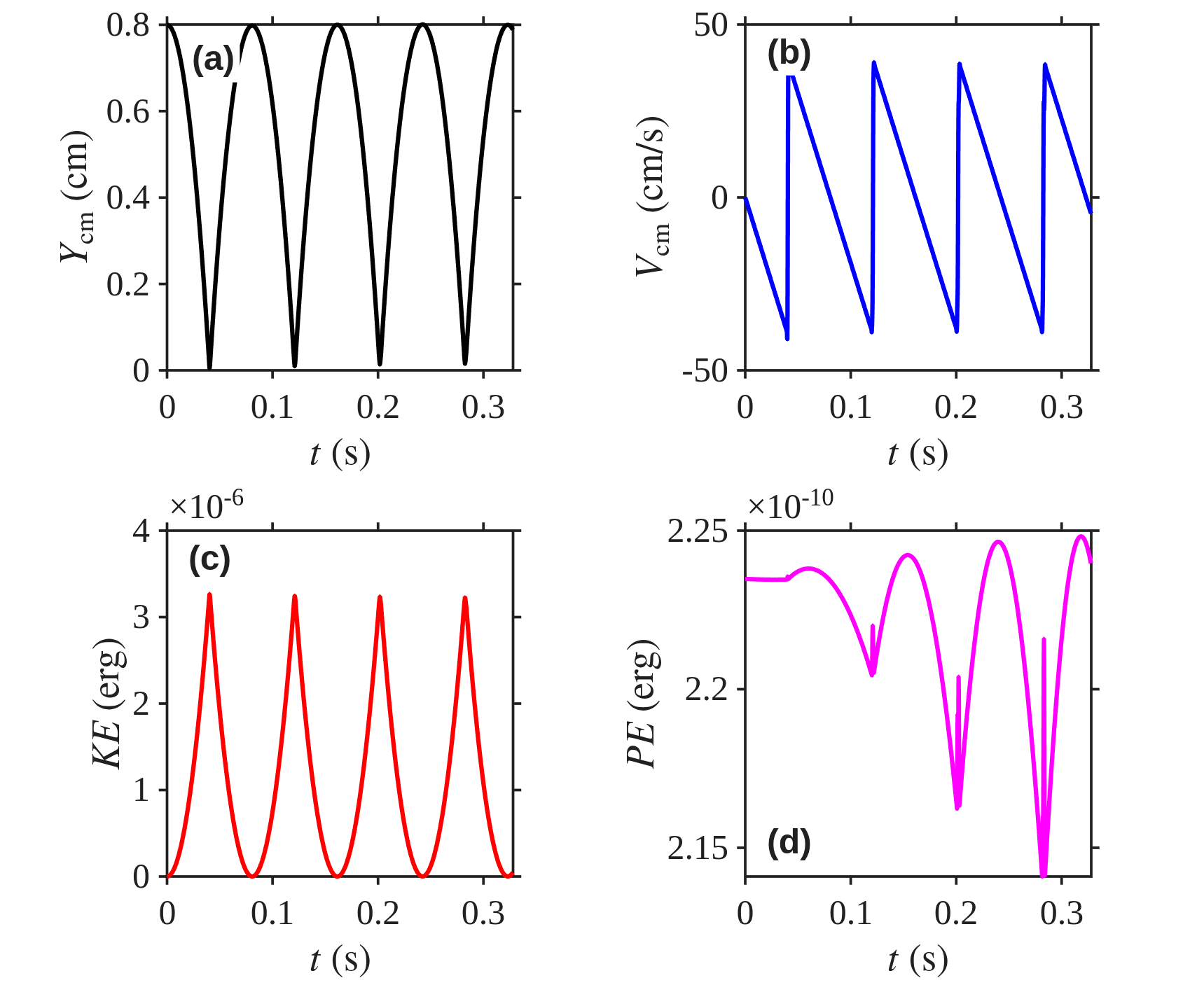}
\caption{
Long-time sedimentation dynamics of a single-layer dusty plasma crystal after removal of the sheath electric field. (a) Center-of-mass height. (b) Center-of-mass vertical velocity. (c) Total kinetic energy. (d) Yukawa potential energy, indicating that the crystal remains structurally intact throughout repeated sedimentation and rebound.
}
\label{fig:figure3}
\end{figure}

The long-time collective dynamics of the crystal are further characterized in Fig.~\ref{fig:figure3}. Figure~\ref{fig:figure3}(a) shows the evolution of the center-of-mass height over four successive sedimentation--rebound cycles. The crystal first reaches the reflecting boundary at $t\approx0.04~\mathrm{s}$, rebounds collectively, and subsequently undergoes repeated collisions at approximately $t\approx0.12$, $0.20$, and $0.28~\mathrm{s}$. Between successive collisions, the crystal rises to its upper turning points at approximately $t\approx0.08$, $0.16$, $0.24$, and $0.32~\mathrm{s}$ before undergoing the next free-fall interval. The complete particle-level evolution throughout these repeated cycles is provided in Supplementary Movie\_S1.

The corresponding center-of-mass velocity, shown in Fig.~\ref{fig:figure3}(b), exhibits the expected piecewise linear variation with slope $-g$ during each free-fall interval and an abrupt sign reversal upon collision with the reflecting boundary, producing nearly identical triangular velocity profiles over successive cycles. This demonstrates that the sedimentation dynamics are highly reproducible, with no measurable change in the gravitational acceleration or rebound characteristics between consecutive collisions, as also observed in Supplementary Movie\_S1.

The kinetic energy shown in Fig.~\ref{fig:figure3}(c) increases continuously during each free-fall interval, reaching maxima immediately before the wall collisions at approximately $t\approx0.04$, $0.12$, $0.20$, and $0.28~\mathrm{s}$. Following each rebound, the kinetic energy decreases to nearly zero near the upper turning points at approximately $t\approx0.08$, $0.16$, $0.24$, and $0.32~\mathrm{s}$ as gravitational potential energy is recovered. The nearly identical kinetic-energy oscillations over successive cycles demonstrate the highly reproducible nature of the sedimentation dynamics. These periodic acceleration and rebound processes are also clearly visualized in Supplementary Movie\_S1.

In contrast, the Yukawa potential energy shown in Fig.~\ref{fig:figure3}(d) remains nearly constant throughout each free-fall interval and exhibits only brief increases during the wall collisions at approximately $t\approx0.04$, $0.12$, $0.20$, and $0.28~\mathrm{s}$, corresponding to the transient compression of the crystal. Following each rebound, the interaction energy rapidly returns to its pre-collision value as the crystal relaxes and resumes collective free fall. The comparatively small and short-lived variation in the interaction energy indicates that the crystal undergoes only modest elastic deformation while preserving its crystalline ordering throughout repeated sedimentation and rebound cycles, consistent with the particle-level dynamics shown in Supplementary Movie\_S1.

\subsection{Sedimentation and Collision Dynamics of a Bilayer Dust Crystal}

To investigate the influence of crystal geometry on the sedimentation dynamics, simulations are performed for a bilayer dusty plasma crystal consisting of 16 identical dust particles arranged in an AB-stacked configuration. As shown in Fig.~\ref{fig:figure1}(b), the increased particle number leads to the spontaneous formation of two hexagonally ordered crystalline layers during equilibration in the presence of the confining electric field and gravity. The equilibrium bilayer consists of two layers, each containing eight dust particles. The lower (blue) and upper (red) layers are centered at approximately $y\approx0.74~\mathrm{cm}$ and $y\approx0.86~\mathrm{cm}$, respectively, corresponding to an equilibrium interlayer spacing of about $0.12~\mathrm{cm}$. The upper layer is laterally shifted relative to the lower layer so that the particles occupy the interstitial sites of the opposite layer, producing the characteristic AB stacking sequence. Both layers retain well-defined in-plane hexagonal ordering with nearly uniform interlayer spacing, indicating that the strong Yukawa coupling stabilizes the bilayer crystal prior to the onset of gravitational sedimentation.

\begin{figure}[ht]
\centering
\includegraphics[width=\linewidth]{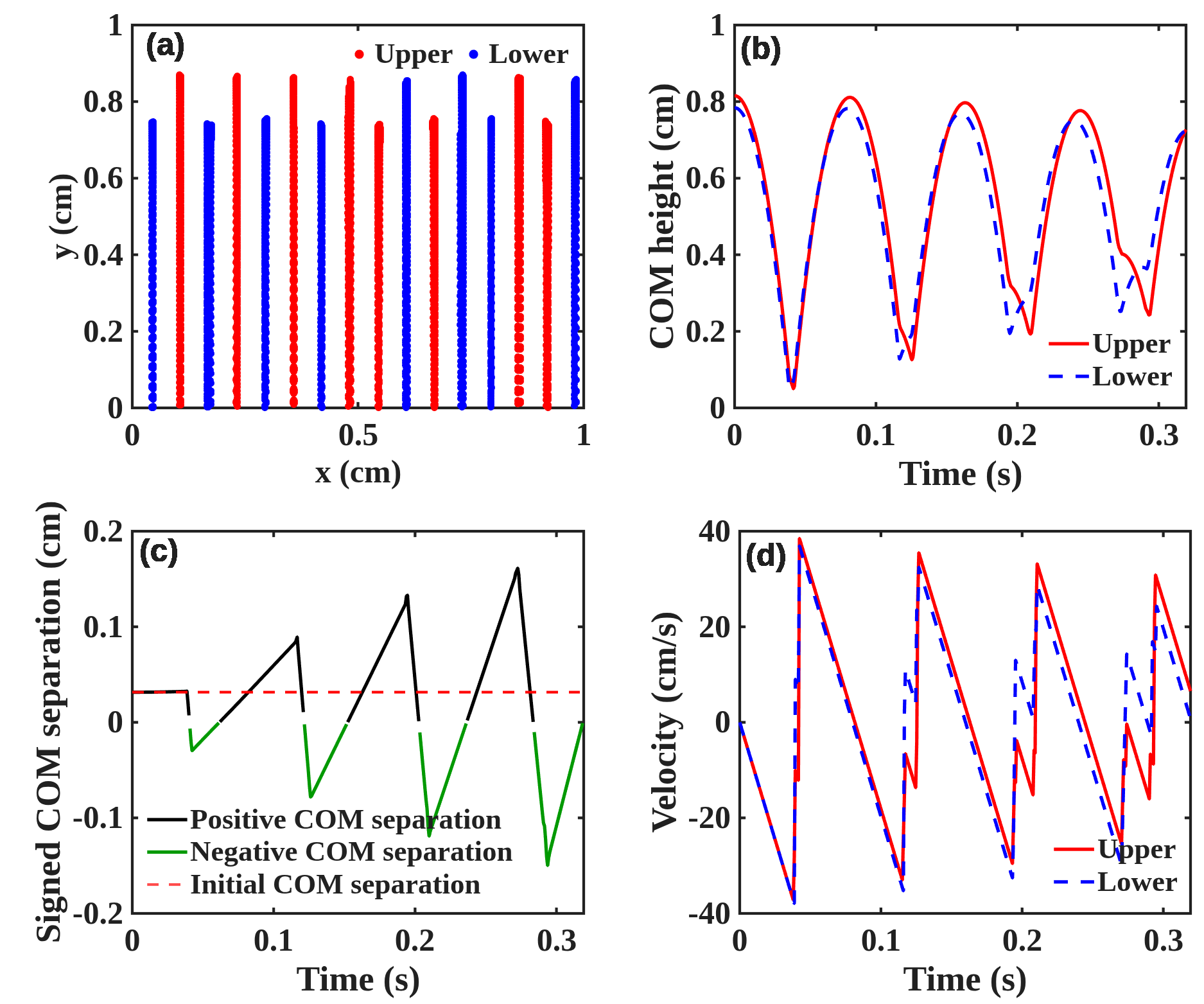}
\caption{
Evolution of a bilayer dusty plasma crystal following removal of the external sheath electric field. (a) Successive snapshots during sedimentation and rebound. (b) Center-of-mass (COM) heights of the upper (red) and lower (blue) layers. (c) Signed COM height difference, showing the transient reversal of the layer ordering during rebound. (d) COM velocities of the two layers, illustrating the delayed reversal of the upper layer due to interlayer Yukawa coupling.
}
\label{fig:figure4}
\end{figure}

Following the removal of the external sheath electric field, the bilayer crystal undergoes collective gravitational sedimentation. Figure~\ref{fig:figure4}(a) presents representative snapshots of the crystal during successive sedimentation, wall collision, and rebound, while Supplementary Movie\_S2 provides a continuous visualization of the complete cyclic dynamics. Initially, the red (upper) layer is located above the blue (lower) layer. During free fall, both layers descend collectively while preserving their in-plane hexagonal ordering and nearly constant interlayer spacing, demonstrating that the bilayer behaves as a mechanically coupled structure. Upon reaching the reflecting boundary, the blue (lower) layer reverses its motion immediately, whereas the red (upper) layer continues to move downward for a short time. This produces a transient compression of the bilayer, during which the rebounding blue layer temporarily rises above the red layer. As momentum is transmitted through the strong interlayer interaction, the compression relaxes, the original layer ordering is recovered, and the bilayer resumes its collective motion. Supplementary Movie\_S2 further shows that this sequence repeats over multiple oscillation cycles.

The corresponding COM dynamics are shown in Fig.~\ref{fig:figure4}(b). During each free-fall interval, the COM heights of the two layers decrease almost in parallel, confirming that the bilayer sediments as a coherent structure. Following each wall collision, however, the COM of the blue layer reverses first, while that of the red layer continues to decrease briefly before rebounding. This transient delay directly reflects the finite time required for momentum to propagate from the lower layer to the upper layer through the interlayer Yukawa interaction. The nearly parallel COM trajectories during successive oscillations demonstrate that the two layers remain strongly coupled throughout the sedimentation process.

Figure~\ref{fig:figure4}(c) quantifies the transient exchange of the vertical positions by plotting the signed COM height difference between the initially identified upper (red) and lower (blue) layers. Positive (black) values indicate that the initially identified upper layer remains above the lower layer, whereas negative (green) values correspond to a temporary reversal of their vertical positions. The horizontal dashed red line denotes the initial equilibrium COM separation between the two layers prior to sedimentation. During the initial free-fall stage, the signed COM separation remains close to this equilibrium value. Following the first wall collision ($t\approx0.04~\mathrm{s}$), the separation becomes slightly negative ($\approx-0.03~\mathrm{cm}$), indicating that the rebounding lower layer temporarily overtakes the upper layer, before increasing to approximately $0.09~\mathrm{cm}$ at $t\approx0.11~\mathrm{s}$. Subsequent collisions at $t\approx0.12~\mathrm{s}$ and $0.21~\mathrm{s}$ produce progressively larger negative excursions of approximately $-0.08$ and $-0.12~\mathrm{cm}$, followed by positive maxima of about $0.13$ and $0.16~\mathrm{cm}$ at $t\approx0.19~\mathrm{s}$ and $0.27~\mathrm{s}$, respectively. After the fourth collision ($t\approx0.29~\mathrm{s}$), the minimum separation further increases in magnitude to approximately $-0.15~\mathrm{cm}$. The repeated crossings of the initial equilibrium line reflect successive transient reversals of the layer positions, while the progressively larger positive and negative excursions indicate increasingly pronounced transient interlayer displacements during repeated rebound cycles. These repeated reversals and the evolution of the signed COM separation are clearly visualized in Supplementary Movie\_S2.

The corresponding COM velocities are shown in Fig.~\ref{fig:figure4}(d). During each free-fall interval, the COM velocities of both layers decrease linearly with time with nearly identical slopes equal to the imposed gravitational acceleration, confirming that the coupled bilayer undergoes uniform gravitational acceleration. Upon collision with the reflecting boundary, however, the velocity of the initially identified lower (blue) layer reverses first, whereas the upper (red) layer continues to accelerate downward for a short interval before reversing. This transient delay demonstrates that momentum is first transferred from the reflecting boundary to the lower layer and is subsequently communicated to the upper layer through the strong interlayer Yukawa interaction. The repeated triangular velocity profiles over successive oscillation cycles indicate that this momentum-transfer mechanism is reproduced after each collision. The delayed velocity reversal and the resulting cyclic exchange of momentum between the two layers are also clearly visualized in Supplementary Movie\_S2. Together, these results demonstrate that the bilayer remains mechanically coherent throughout repeated sedimentation and rebound cycles despite repeated collisions with the reflecting boundary.

\subsection{Sedimentation and Collision Dynamics of a Trilayer Dust Crystal}

To further investigate the influence of crystal thickness on the sedimentation dynamics, the number of dust particles is increased from 16 in the bilayer configuration to 24 while all other simulation parameters, including the Yukawa interaction strength, gravitational acceleration, the confining electric field, and the confinement geometry, are kept unchanged. As shown in Fig.~\ref{fig:figure1}(c), the particles spontaneously self-organize during equilibration into an ABA-stacked trilayer crystal consisting of three nearly hexagonal layers, each containing eight dust particles. The lower, middle, and upper layers are centered at approximately $y\approx0.70~\mathrm{cm}$, $0.80~\mathrm{cm}$, and $0.90~\mathrm{cm}$, respectively, giving nearly uniform interlayer spacings of about $0.10~\mathrm{cm}$. The middle layer is laterally shifted relative to the upper and lower layers, producing the characteristic ABA stacking sequence. The well-defined hexagonal ordering and uniform interlayer spacing demonstrate that the strong Yukawa interactions stabilize the equilibrium trilayer crystal prior to gravitational sedimentation.

Compared with the bilayer crystal, the additional layer introduces a second interlayer interface, increasing the degree of mechanical coupling within the system. This provides an opportunity to investigate how momentum propagates through multiple interfaces and influences the subsequent sedimentation and rebound dynamics.

Following the removal of the confining electric field, the trilayer crystal undergoes collective gravitational sedimentation. During free fall, the three layers descend together while preserving their in-plane hexagonal ordering and nearly constant interlayer spacing, demonstrating that the crystal behaves as a mechanically coupled structure rather than as three independently falling layers. As the crystal reaches the reflecting boundary, the lower layer collides with the wall first and reverses its vertical motion, while the middle and upper layers continue to move downward for a short interval. The reflected momentum is subsequently transmitted from the lower layer to the middle layer and finally to the upper layer through the strong interlayer Yukawa interactions.

\begin{figure}[ht]
\centering
\includegraphics[width=\linewidth]{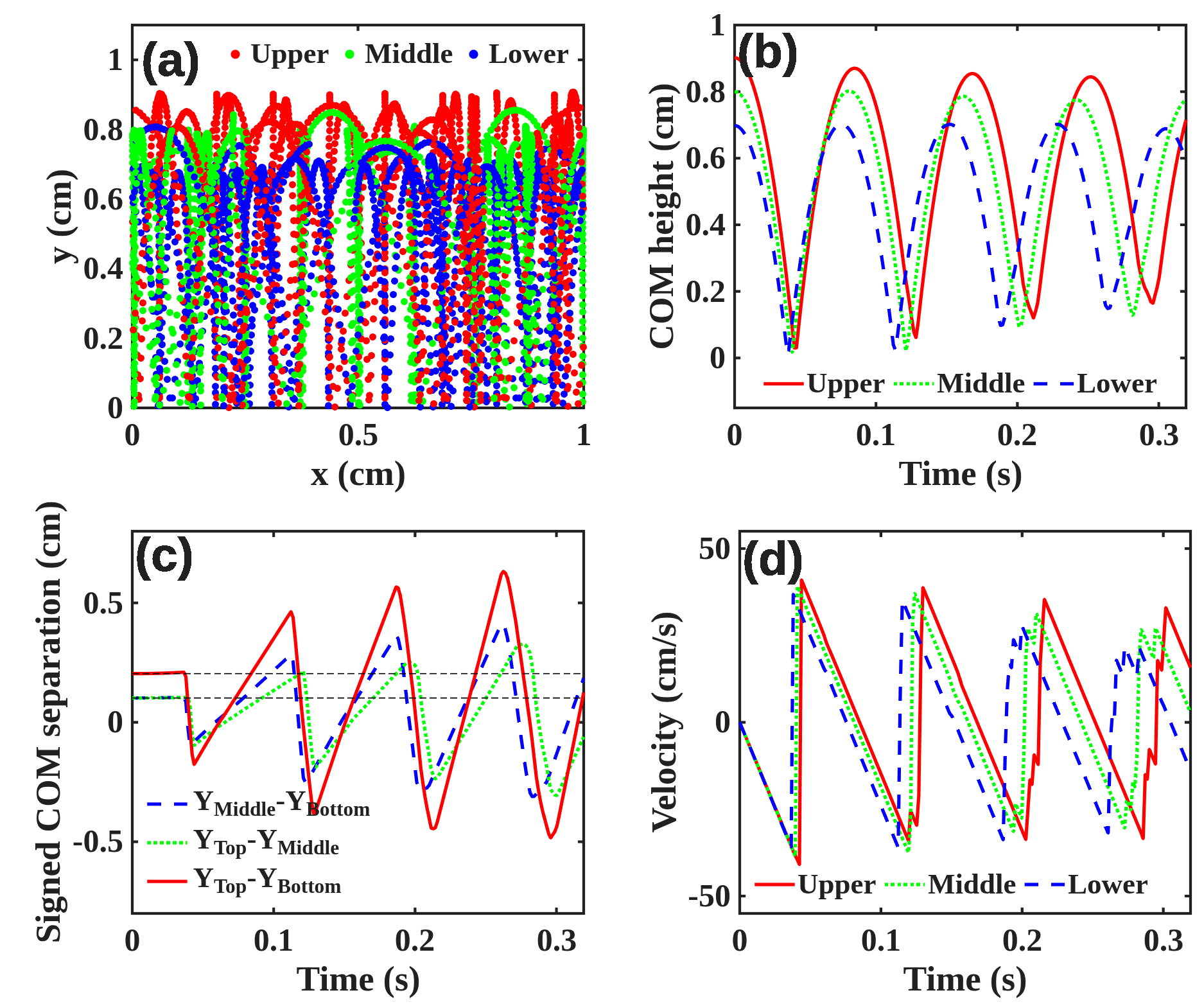}
\caption{
Evolution of a trilayer dusty plasma crystal following removal of the external sheath electric field. (a) Successive snapshots during sedimentation and rebound. (b) Center-of-mass heights of the three layers. (c) Signed center-of-mass height differences; dashed lines indicate the equilibrium separations and the dotted line denotes zero. (d) Center-of-mass velocities, showing the sequential propagation of momentum through the trilayer.
}
\label{fig:figure5}
\end{figure}

Figure~\ref{fig:figure5}(a) presents representative snapshots of the trilayer during repeated gravitational sedimentation, wall collision, and rebound. At $t=0$, the crystal forms a well-ordered ABA-stacked structure, where the lower and upper A-type layers are shown in blue and red, respectively, while the middle B-type layer is shown in green, with nearly uniform interlayer spacing. During the initial free-fall stage, the three layers descend collectively while preserving their in-plane hexagonal ordering, indicating that the crystal behaves as a mechanically coupled structure. When the lower blue A-type layer reaches the reflecting boundary at approximately $t\approx0.04~\mathrm{s}$, it immediately reverses direction, whereas the middle green B-type layer and the upper red A-type layer continue moving downward. Between $t\approx0.0401$ and $0.0408~\mathrm{s}$, the rebounding lower blue A-type layer overlaps with the descending upper red A-type layer and interacts through the interlayer Yukawa force while exchanging their relative vertical positions. At approximately $t\approx0.0405~\mathrm{s}$, the lower blue layer temporarily overlaps and appears to cover the upper red layer in the projection. Following this interaction, the lower blue A-type layer continues upward while the upper red A-type layer continues downward, producing a slight distortion of the initial ABA layer registry without disrupting the collective motion of the trilayer. During subsequent sedimentation--rebound cycles at approximately $t\approx0.12$, $0.20$, and $0.28~\mathrm{s}$, repeated interlayer interactions progressively degrade the initial ABA stacking, leading to an increasing degree of structural disorder. Nevertheless, the trilayer remains mechanically coherent and continues to undergo collective sedimentation and rebound owing to the strong interlayer Yukawa coupling. The complete particle-level evolution throughout these repeated cycles is shown in Supplementary Movie\_S3.

The corresponding COM heights are shown in Fig.~\ref{fig:figure5}(b). During each free-fall interval, the COM trajectories of the three layers decrease nearly in parallel, demonstrating coherent collective motion under gravity. Following each collision with the reflecting boundary, however, the lower, middle, and upper layers reverse sequentially, directly revealing the finite propagation time of momentum through the strongly coupled trilayer. This sequential response is reproduced over successive oscillation cycles, indicating that the three layers remain dynamically coupled throughout the evolution. Although the COM trajectories remain highly coherent, the particle configurations shown in Fig.~\ref{fig:figure5}(a) and Supplementary Movie\_S3 reveal a gradual degradation of the initial ABA layer registry while the collective oscillatory motion is preserved.

The transient evolution of the interlayer separations is presented in Fig.~\ref{fig:figure5}(c), which shows the signed COM height differences between the middle and lower layers, the upper and middle layers, and the upper and lower layers. During the initial free-fall stage, these quantities remain close to their equilibrium values, confirming that the trilayer sediments as a mechanically coupled structure. Following the first wall collision at approximately $t\approx0.04~\mathrm{s}$, the signed COM separations undergo transient sign reversals as the lower and middle layers rebound while the upper layer continues its downward motion. During the subsequent rebound, the separations recover and temporarily exceed their initial equilibrium values before the onset of the next free-fall interval. The same sequence repeats following the collisions at approximately $t\approx0.12$, $0.20$, and $0.28~\mathrm{s}$, with progressively larger positive and negative excursions. These increasing excursions indicate progressively larger transient deviations from the initial layer registry, reflecting the cumulative effect of repeated interlayer interactions observed in Supplementary Movie\_S3.

The corresponding COM velocities are presented in Fig.~\ref{fig:figure5}(d). During each free-fall interval, the velocities of all three layers decrease linearly with time with nearly identical slopes equal to the imposed gravitational acceleration. At each wall collision, the lower layer undergoes the earliest velocity reversal, followed successively by the middle and upper layers after short delays, demonstrating the propagation of the wall-induced impulse through the strongly coupled trilayer. As the number of oscillation cycles increases, the velocity profiles become progressively more irregular owing to repeated interlayer momentum exchange and the increasingly complex particle dynamics that develop after successive rebounds. Nevertheless, the three layers continue to oscillate collectively, demonstrating that strong interlayer Yukawa coupling preserves the overall mechanical coherence of the trilayer despite the gradual degradation of the initial ABA stacking, as clearly visualized in Supplementary Movie\_S3.

The observed sequential momentum propagation resembles the transmission of a mechanical disturbance through an elastically coupled lattice. In the present system, however, the coupling does not arise from direct contact forces between neighboring particles but from long-range Yukawa interactions. Consequently, the wall-induced impulse propagates through the multilayer crystal via collective many-body interactions, giving rise to the sequential layer-by-layer velocity reversal shown in Fig.~\ref{fig:figure5}(d).

\section{Conclusions and Outlook} 
\label{Conclusions} 

In this work, we have investigated the gravitational sedimentation and rebound dynamics of strongly coupled dusty plasma crystals using molecular dynamics simulations. Following the removal of electrostatic levitation, single-layer, AB-stacked bilayer, and ABA-stacked trilayer crystals undergo collective gravitational sedimentation while interacting through the Yukawa (screened Coulomb) potential. By systematically comparing these crystal configurations, we have demonstrated that crystal geometry plays a crucial role in determining the subsequent collision and rebound dynamics.

The simulations reveal that strong Yukawa interactions preserve the collective mechanical response of the crystal throughout repeated sedimentation--rebound cycles despite the continual action of gravity. While single-layer crystals exhibit coherent oscillatory motion following collision with the reflecting boundary, multilayer crystals display considerably richer dynamics arising from interlayer coupling. In particular, impact with the boundary generates transient interlayer compression followed by sequential momentum transfer between neighboring layers, producing delayed rebound of the upper layers. For the ABA-stacked trilayer, repeated collisions progressively degrade the initial stacking sequence through layer exchange while maintaining coherent collective oscillations. These results demonstrate that strongly coupled dusty plasma crystals behave as mechanically resilient many-body systems capable of sustaining repeated impacts without losing their collective dynamical character.

Beyond elucidating the sedimentation dynamics of dusty plasma crystals, the present study provides a particle-resolved picture of momentum propagation and structural evolution in strongly coupled Yukawa systems. The observed layer-by-layer transmission of momentum and the gradual evolution of crystal stacking highlight the important role of crystal geometry in governing collective impact dynamics. Although the present model is idealized, with constant dust charge, fixed Yukawa interactions, and a reflecting boundary, it isolates the fundamental mechanical response of strongly coupled plasma crystals and provides a useful reference for future theoretical, numerical, and laboratory investigations.

The underlying physical mechanisms identified here are also relevant to other plasma environments involving charged particulate transport. In plasma-assisted semiconductor manufacturing, for example, charged particles generated within plasma reactors may become trapped in plasma sheaths and subsequently settle onto wafer or chamber surfaces following changes in plasma operating conditions. Although the present simulations do not model such systems directly, they provide fundamental physical insight into how strong interparticle interactions influence collective sedimentation, momentum transfer, and particle--surface impact dynamics.

Future work will extend the present model by incorporating neutral gas drag, realistic sheath electric fields, time-dependent particle charging, plasma flow, external electric and magnetic fields, larger multilayer crystals, binary dusty plasmas, and absorbing boundaries representative of practical plasma processing environments. Such studies will improve the realism of the model while advancing the understanding of collective transport, impact dynamics, and structural evolution in strongly coupled dusty plasmas.

\section{Acknowledgments}
GG gratefully acknowledges the financial support provided by St. Xavier's College (Autonomous), Ahmedabad, through the Xavier's Minor Research Project Seed Grant (SXCA/XIP/2025–2026/XMRP04).
\section*{Author Declaration}
\subsection*{Conflict of interest}
The authors report no conflict of interest.	

\section*{Data Availability}

The data that support the findings of this study are available from the corresponding author upon reasonable request.

\bibliographystyle{unsrt}
\bibliography{vikram_dst}
	
\end{document}